\documentclass[]{pasj02} 
\usepackage[switch,mathlines]{lineno} 
\usepackage{booktabs}
\usepackage{url}
\usepackage{multirow}
\usepackage{natbib}

\jyear{2024}
\Received{}
\Accepted{}

\begin{document} 

\title{XRISM view of a stellar flare: High-resolution Fe K spectra of HR~1099, an RS CVn-type star}

\author{Miki~\textsc{Kurihara}\altaffilmark{1,2}\altemailmark\orcid{0000-0002-3133-9053}\email{kurihara@ac.jaxa.jp, miki39kurihara@g.ecc.u-tokyo.ac.jp}} 
\author{Masahiro~\textsc{Tsujimoto}\altaffilmark{1,2}\orcid{0000-0002-9184-5556}}
\author{Michael~\textsc{Loewenstein}\altaffilmark{3}\orcid{0000-0002-1661-4029}}
\author{Yoshitomo~\textsc{Maeda}\altaffilmark{2}\orcid{0000-0002-9099-5755}}
\author{Marc~\textsc{Audard}\altaffilmark{4}\orcid{0000-0003-4721-034X}}
\author{Ehud~\textsc{Behar}\altaffilmark{5}\orcid{0000-0001-9735-4873}}
\author{Megan~E.~\textsc{Eckart}\altaffilmark{6}\orcid{0000-0003-3894-5889}}
\author{Adam~\textsc{Foster}\altaffilmark{7}\orcid{0000-0003-3462-8886}}
\author{Liyi~\textsc{Gu}\altaffilmark{8}\orcid{0000-0001-9911-7038}}
\author{Matteo~\textsc{Guainazzi}\altaffilmark{9}\orcid{0000-0002-1094-3147}}
\author{Kenji~\textsc{Hamaguchi}\altaffilmark{10,3,11}\orcid{0000-0001-7515-2779}}
\author{Natalie~\textsc{Hell}\altaffilmark{6}\orcid{0000-0003-3057-1536}}
\author{Shun~\textsc{Inoue}\altaffilmark{12}\orcid{0000-0003-3085-304X}}
\author{Yukiko~\textsc{Ishihara}\altaffilmark{13}}
\author{Satoru~\textsc{Katsuda}\altaffilmark{14}\orcid{0000-0002-1104-7205}}
\author{Caroline~A.~\textsc{Kilbourne}\altaffilmark{3}\orcid{0000-0001-9464-4103}}
\author{Maurice~A.~\textsc{Leutenegger}\altaffilmark{3}\orcid{0000-0002-3331-7595}}
\author{Eric~D.~\textsc{Miller}\altaffilmark{15}\orcid{0000-0002-3031-2326}}
\author{Nagisa~\textsc{Nagashima}\altaffilmark{13}}
\author{Frederick~Scott~\textsc{Porter}\altaffilmark{3}\orcid{0000-0002-6374-1119}}
\author{Makoto~\textsc{Sawada}\altaffilmark{16}\orcid{0000-0003-2008-6887}}
\author{Yohko~\textsc{Tsuboi}\altaffilmark{13}\orcid{0000-0001-9943-0024}}
\author{Vinay~L.~\textsc{Kashyap}\altaffilmark{7}\orcid{0000-0002-3869-7996}}
\author{David~H.~\textsc{Brooks}\altaffilmark{17,18,2}\orcid{0000-0002-2189-9313}}

\altaffiltext{1}{Department of Astronomy, Graduate School of Science, The University of Tokyo, Bunkyo-ku, Tokyo 113-0033, Japan}
\altaffiltext{2}{Japan Aerospace Exploration Agency, Institute of Space and Astronautical Science, Chuo-ku, Sagamihara, Kanagawa 252-5210, Japan}
\altaffiltext{3}{NASA / Goddard Space Flight Center, Greenbelt, MD 20771, USA}
\altaffiltext{4}{Department of Astronomy, University of Geneva, Versoix CH-1290, Switzerland}
\altaffiltext{5}{Department of Physics, Technion, Technion City, Haifa 3200003, Israel}
\altaffiltext{6}{Lawrence Livermore National Laboratory, Livermore, CA 94550, USA}
\altaffiltext{7}{Center for Astrophysics, Harvard-Smithsonian, Cambridge, MA 02138, USA}
\altaffiltext{8}{SRON Netherlands Institute for Space Research, Leiden, The Netherlands}
\altaffiltext{9}{European Space Agency, European Space Research and Technology Centre, Noordwijk, The Netherlands}
\altaffiltext{10}{Center for Space Science and Technology, University of Maryland, Baltimore County, Baltimore, MD 21250, USA}
\altaffiltext{11}{Center for Research and Exploration in Space Science and Technology (CRESST) II, NASA / Goddard Space Flight Center, Greenbelt, MD 20771, USA}
\altaffiltext{12}{Department of Physics, Kyoto University, Sakyo-ku, Kyoto, Kyoto 606-8502, Japan}
\altaffiltext{13}{Department of Physics, Chuo University, Bunkyo-ku, Tokyo 112-8551, Japan}
\altaffiltext{14}{Department of Physics, Saitama University, Sakura-ku, Saitama, Saitama 338-8570, Japan}
\altaffiltext{15}{Kavli Institute for Astrophysics and Space Research, Massachusetts Institute of Technology, Cambridge, MA 02139, USA}
\altaffiltext{16}{Department of Physics, Rikkyo University, Toshima-ku, Tokyo 171-8501, Japan} 
\altaffiltext{17}{Department of Physics \& Astronomy, George Mason University, 4400 University Drive, Fairfax, VA 22030, USA}
\altaffiltext{18}{University College London, Mullard Space Science Laboratory, Dorking, Surrey, RH5 6NT, UK}

\maketitle

\begin{abstract}
 A high-resolution X-ray spectroscopic observation was made of the RS CVn-type binary
 star HR~1099 using the Resolve instrument onboard XRISM for its calibration
 purposes. During the $\sim$400~ks telescope time covering 1.5 binary orbit, a flare
 lasting for $\sim$100~ks was observed with a released X-ray radiation energy of $\sim
 10^{34}$~erg, making it the first stellar flare ever observed with an X-ray
 microcalorimeter spectrometer. The flare peak count rate is 6.4 times higher than that
 in quiescence and is distinguished clearly in time thanks to the long telescope
 time. Many emission lines were detected in the 1.7--10~keV range both in the flare and
 quiescent phases. Using the high spectral resolution of Resolve in the Fe K band
 (6.5--7.0~keV), we resolved the inner-shell lines of Fe
 \emissiontype{XIX}--\emissiontype{XXIV} as well as the outer-shell lines of Fe
 \emissiontype{XXV}--\emissiontype{XXVI}. These lines have peaks in the contribution
 functions at different temperatures over a wide range, allowing us to construct the
 differential emission measure (DEM) distribution over the electron temperature of
 1--10~keV (roughly 10--100~MK) based only on Fe lines, thus without an assumption of
 the elemental abundance. The reconstructed DEM has a bimodal distribution, and only the
 hotter component increased during the flare. The elemental abundance was derived based
 on the DEM distribution thus constructed.  A significant abundance increase was
 observed during the flare for Ca and Fe, which are some of the elements with the lowest
 first ionization potential among those analyzed, but not for Si, S, and Ar.  This
 behavior is seen in some giant solar flares and the present result is a clear example
 in stellar flares.
\end{abstract}

\section{Introduction}\label{s0}
Flares on stars are among the most energetic manifestations of magnetic activity and
continue to be intensively studied (e.g., \citealt{benz2010,kowalski2024}). The standard
paradigm of eruptive flares suggests that magnetic reconnection impulsively releases the
energy stored in stellar coronae \citep{priest2002,shibata2011}, producing optically
thin plasma at temperatures of $\sim$0.1--10~keV that emits both thermal and nonthermal
radiation in X-rays. Key properties of coronal plasmas include their thermal structure
and chemical composition, as these directly relate to the heating and cooling
processes. The physics is considered to be the same with solar flares but observations
of magnetically active stars have revealed phenomena not typically seen in the Sun, such
as more energetic flares (e.g., \citealt{tsuboi2016}) and the inverse first ionization
potential (FIP) effect (see \citealt{laming2009,laming2015} for reviews).

Since the first discovery of abundance differences between the solar photosphere and
corona by \citet{pottasch1963}, studies of the FIP effect have expanded to distant stars
(\citealt{testa2010} for a review). The FIP effect refers to the systematic enrichment
of low-FIP elements ($\lesssim 10$ eV) or the depletion of high-FIP elements in stellar coronae relative to their photospheric
abundances, while the inverse FIP (i-FIP) effect denotes the opposite trend.  Although the physical
origin remains debated, these behaviors are thought to be related to the coronal
heating mechanism itself. Comparative analyses across stellar samples indicate that the
strength and sign of the FIP bias depend on activity level \citep{telleschi2005,seli2022}, and this
has been extensively studied using X-ray grating spectrometers (e.g.,
\citealt{audard2003a,nordon2008}). The target of this study, the RS CVn-type binary HR~1099, is a
representative system that served as the first-light target of XMM-Newton
\citep{audard2001} and is well known to exhibit the i-FIP effect
\citep{brinkman2001,drake2001,nordon2008,seli2022}.

\medskip

In this paper, we present the results of a week-long Resolve calibration
observation of HR~1099. A stellar flare was captured for the first
time with an X-ray microcalorimeter spectrometer. Using this data set, we perform the
differential emission measure (DEM) analysis for both the flaring and quiescent phases using the Fe
\emissiontype{XIX}--\emissiontype{XXVI} K-shell lines, based on which we investigate the
changes of the chemical abundance between the two phases.

The outline of this paper is as follows. In \S~\ref{s1}, we first describe the method of
using the Fe \emissiontype{XIX}--\emissiontype{XXVI} K-shell lines ratios for plasma
diagnostics. In \S~\ref{s2}, we describe the target
(\S~\ref{s2-1}), the observation (\S~\ref{s2-2}), and the data reduction
(\S~\ref{s2-3}). Some technical details are given to explain the context of this
calibration observation made in unusual instrumental configurations and data processing
specific to the data. In \S~\ref{s3}, we present the Resolve light curve and spectra
(\S~\ref{s3-1}), followed by the results of the spectral modeling in the broadband
(\S~\ref{s3-2}) and the Fe K complexes (\S~\ref{s3-3}). In \S~\ref{s4}, we present two
results; the DEM distribution (\S~\ref{s4-1}) and the chemical abundance
(\S~\ref{s4-2}), and compare the present data with some previous results. In
\S~\ref{s5}, we conclude with major findings of this study.

\section{Method}\label{s1}
\begin{figure*}[!bthp]
 \begin{center}
 \includegraphics[width=0.99\linewidth]{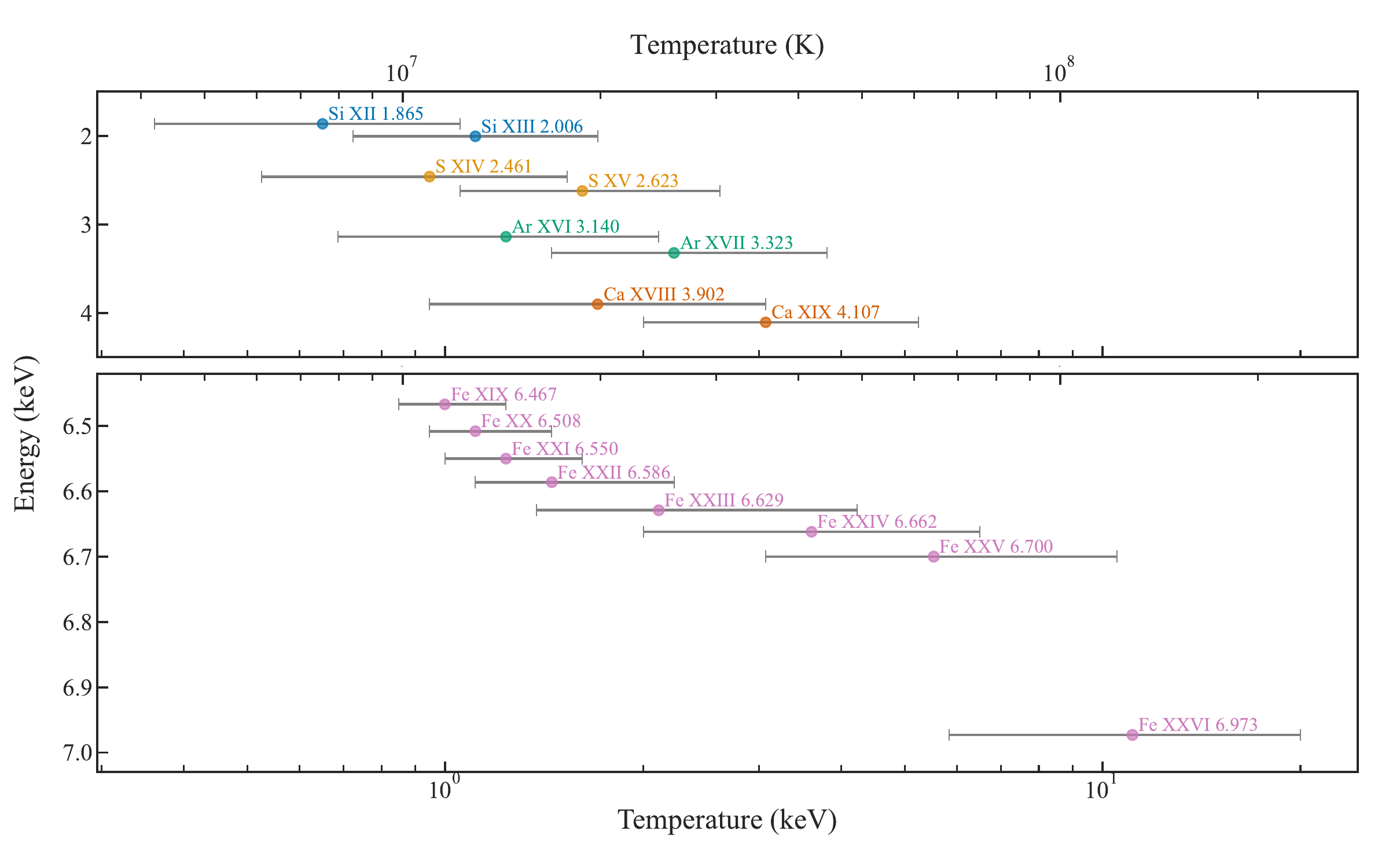}
 \end{center}
 \caption{Temperature range at which representative emission lines of different elements
 in different charge states are formed. Their ion name and the line energy in keV are
 given for each dot. The dots are placed at the peak of the contribution function (at
 the low density limit of $10^{10}$~cm$^{-3}$) on the horizontal axis and the line
 energy on the vertical axis. The horizontal bars indicate the temperature range where
 the contribution function exceeds half of the peak value. For Fe, selected lines of
 the lower charge states (Fe \emissiontype{XIX}--\emissiontype{XXII}) are predominantly
 formed by the dielectronic recombination process, while those of the higher charge
 states (Fe \emissiontype{III}--\emissiontype{XXVI}) are by direct excitation.  For Si,
 S, Ar, and Ca, the $w$ line of the He-like and the Ly$\alpha1$ line of the H-like ions
 are selected. \texttt{Chianti} version 11 \citep{dere1997,dufresne2024} is used.
 Alt text: A figure showing temperature ranges that Si, S, Ar, Ca, and Fe lines can cover.
 }
 \label{f03}
\end{figure*}

In distant stars, neither spatially resolved imaging nor in-situ measurements are
possible. Spectroscopy therefore provides the primary means to diagnose the thermal
structure and chemical composition of coronal plasma. The thermal structure is commonly
characterized by DEM distributions, which represent the emission measure of plasma at
each temperature bin integrated over the stellar disk. DEMs for solar and stellar
coronae have been constructed from emission lines in EUV and X-ray spectra. Several
inversion techniques for DEM reconstruction have been developed for the Sun (e.g.,
\citealt{del2018} for a review). Emission line features originate from bound-bound
transitions of various elements in different charge states that are affected by thermal
structure. A key point here is to use emission lines from ions of a single element
across a wide range of ionization states, providing broad temperature coverage and
decoupling the degeneracy between DEM shape and elemental abundances.

In stellar X-ray studies, this strategy became feasible with the advent of the
grating spectrometers onboard the Chandra
\citep{weisskopf2000} and XMM-Newton \citep{jansen2001} X-ray observatories. In
practice, Fe is the only element that can be used for this purpose. Using the Reflection
Grating Spectrometer (RGS; \citealt{den2001}), \citet{brinkman2001} used a series of Fe
\emissiontype{XVII}--\emissiontype{XXIV} L-shell lines at $\sim$1~keV. Using the High
Energy Transmission Grating (HETG; \citealt{canizares2005}), \citet{nordon2008} used the
Fe \emissiontype{XVII}--\emissiontype{XXIV} L-shell lines as well as the Fe
\emissiontype{XXV} K-shell line at 6.7~keV. 
However, systematic uncertainties remain in the use of this Fe \emissiontype{XXV} K-shell line in grating missions as unresolved contributions from the lower charged Fe K-shell lines persist.
Also, the Fe
\emissiontype{XXVI} lines should be added to constrain the hottest end of the DEM
distribution.

In this regard, the use of Fe K-shell lines in the 6.4--7.0~keV range is particularly
advantageous. This energy range contains outer-shell transition lines of Fe \emissiontype{XXVI}
and \emissiontype{XXV} as well as inner-shell transition lines of Fe \emissiontype{XXIV}
and lower charge states, which have become possible with the advent of the
microcalorimeter spectrometer. 
The Resolve instrument \citep{kelley2025,ishisaki2022} onboard the X-ray
Imaging and Spectroscopy Mission (XRISM; \citealt{tashiro2020}) provides high-resolution
spectra ($R \equiv E/\Delta E \sim 1300$ at the Fe K band) along with broad energy
coverage from 1.7 to 12.0~keV. The Fe K-shell lines from different charge states can be
resolved for constructing the DEM distribution over a wide temperature range. In
addition, strong K-shell lines of Si through Ni can be used to derive the elemental abundance.

Figure~\ref{f03} illustrates the concept of this diagnostic. The temperature range, in
which representative lines of different charge states are dominant, are shown. For
example, the Fe\emissiontype{XXVI} Ly$\alpha$ lines (2p $^{2}P$ $\rightarrow$ 1s $^{2}S$)
at 7.0~keV are formed primarily by the direct excitation in H-like Fe ion at the ground
state followed by radiative decay. Therefore, the contribution function of the line
reflects H-like Fe population function dominant over a temperature range of
10$^{0.4}$--10$^{1.1}$~keV. Other lines cover different ranges. When combined, the Fe
K-shell lines in the 6.4--7.0~keV range alone can constrain the DEM distribution in a
broad range of $\gtrsim 1$~dex. In the same temperature range, H-like and He-like Si, S,
Ar, and Ca K-shell lines can be easily measured and used for constraining the elemental
abundance across a wide FIP range from the low (Ca) to the high (Ar) end.

\begin{figure*}[!hbtp]
 \begin{center}
 \includegraphics[width=0.99\linewidth]{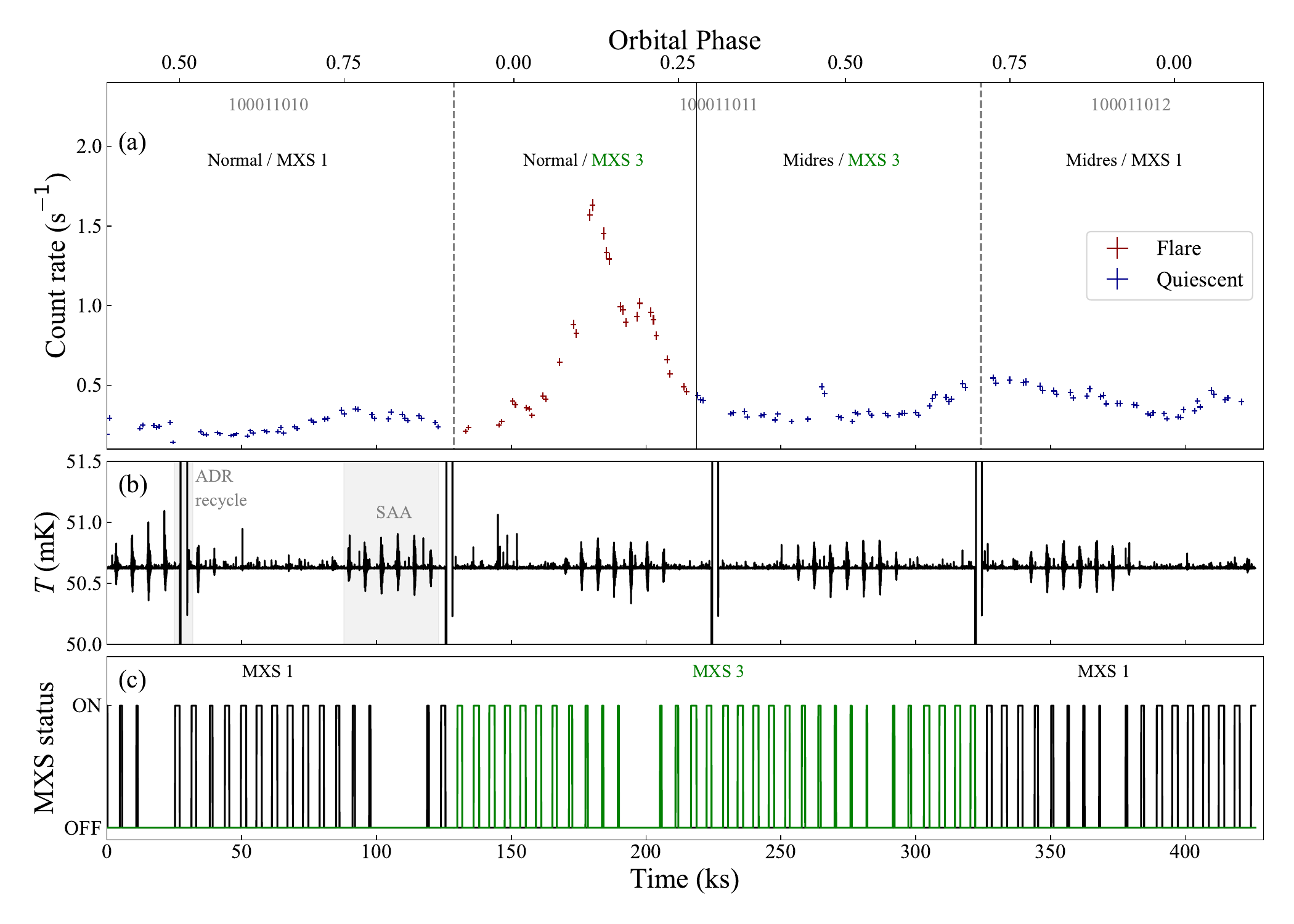}
 \end{center}
 \caption{(a) Resolve light curve of HR 1099 in the 1.7--10.0 keV with a 1024~s
 binning after good-time-interval screening. The reference time is
 2024-03-06T01:15:32. The binary phase (i.e., the orbital phase of the binary, with the primary star being in front at phase zero) is shown on the top axis. The modes of the
 observation are shown for the four parts. The flare phase was chosen to be the same
 with the second part, as it encompasses most of the flare. 
 (b) Temperature of the 50~mK stage of the detector cooling system. The four large deviations at around 25, 125, 225, and
 325~ks are due to ADR recycles. Other repeated deviations are mostly due to spacecraft
 passages of the South Atlantic Anomaly (SAA) region.
 (c) Status of the MXS on or off. MXS1 is in black, while MXS3 is in green. They were
 turned on when the source was eclipsed by the Earth. The $^{55}$Fe sources on the
 filter wheel were also rotated into the aperture during the same intervals.  Alt text:
 A three-panel figure showing (a) X-ray light curve in the 1.7–-10.0 keV energy range,
 (b) 50~mK stage temperature, and (c) MXS status during the observation.}
 \label{f01}
\end{figure*}

\begin{figure*}[!hbtp]
 \begin{center}
 \includegraphics[width=0.99\linewidth]{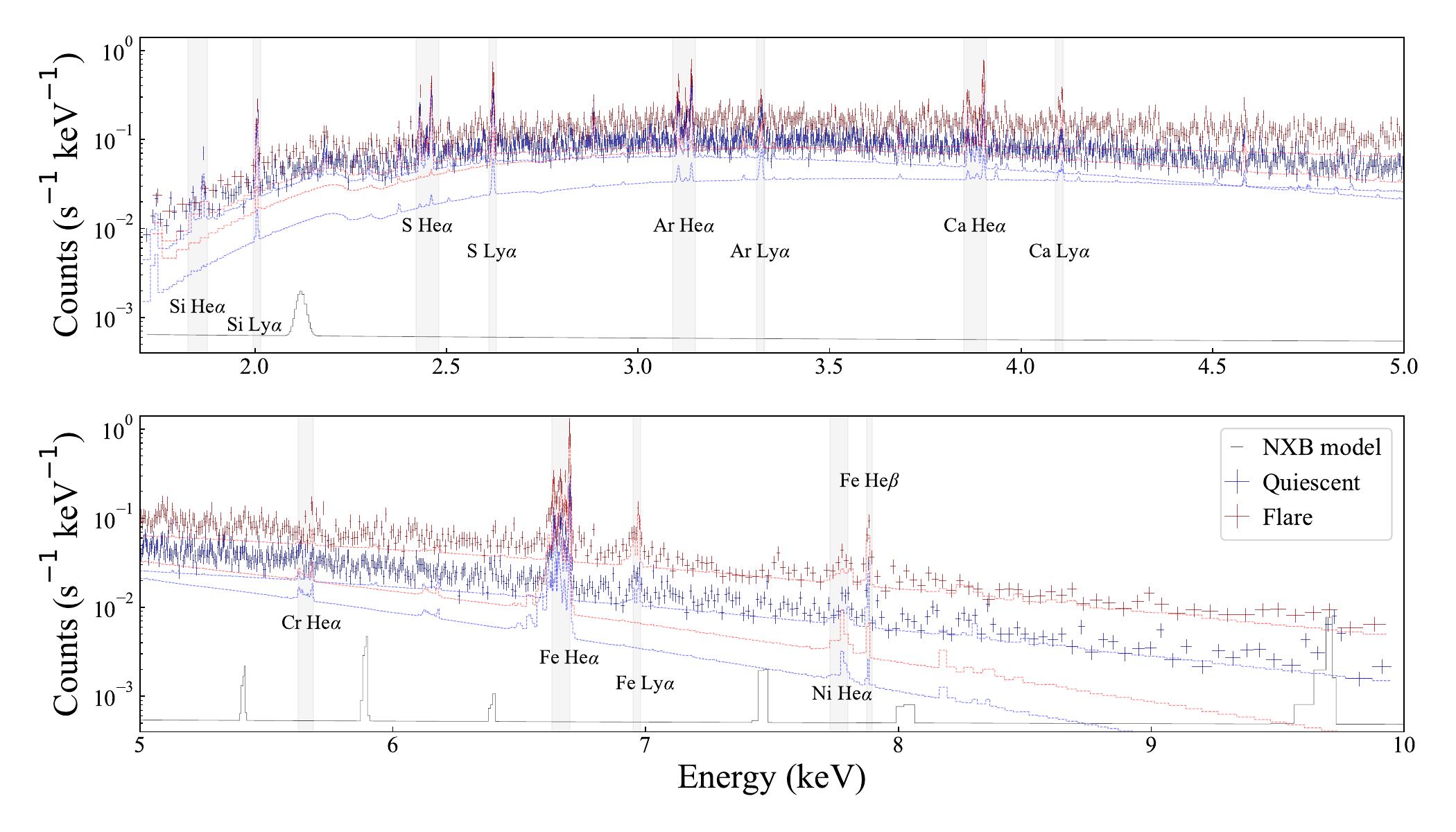}
 \end{center}
 \caption{Resolve spectrum in the 1.7--10.0 keV for the
 flaring (red) and quiescent data (blue) phase as well as the background model spectrum
 (black). The best-fit model of the fiducial 2$kT$ fit components is also shown. The gray stripes indicate the energy range of the conspicuous He$\alpha$
 and Ly$\alpha$ lines. The data were binned to have a minimum of 30 events per bin for this figure.
 Alt text: Resolve X-ray spectrum in the 1.7--10.0 keV range. 
 }
 \label{f02}
\end{figure*}

\section{Target and observation}\label{s2}
\subsection{Target}\label{s2-1}
The observation target is an RS CVn type binary, HR~1099 (V711 Tau). It is located at a
distance of $\sim$29.4 pc and consists of two sub-giants: K1 IV primary and G5 IV--V
secondary \citep{fekel1983} with its orbital period of $P=2.837711\pm0.000066$ day and
an epoch of the phase origin at HJD$=2457729.7084\pm0.0017$ \citep{strassmeier2020}. The
radii of the primary and secondary stars are estimated to be $3.3-4.3~R_\odot$ and
$1.1-1.5~R_\odot$, respectively, for the assumed inclination angle $30-40^{\circ}$
\citep{fekel1983}. The radial velocities with respect to the Earth have an amplitude of
$K_1 = 48.47~\mathrm{km~s^{-1}}$ (equivalent Doppler shift of 1.4~eV at 6.7~keV) and
$K_2=62.41~\mathrm{km~s^{-1}}$ (1.1~eV) with a systemic velocity of
$-14.76~\mathrm{km~s^{-1}}$ (--0.33~eV).

Leveraging its proximity and brightness, extensive multi-wavelength observations have
been conducted for HR~1099. The primary star is the more active, exhibiting strong
chromospheric and coronal activities \citep{golay2024}. The existence of star spots is
also indicated based on Zeeman Doppler imaging (e.g., \citealt{petit2004}) and
photometric observations (e.g., \citealt{berdyugina2007}).

The star is also active in the X-rays \citep{perdelwitz2018}, which is known to be
predominantly from the K1 IV primary \citep{ayres2001}. Many high-resolution X-ray
spectroscopic observations were made using the grating spectrometers onboard XMM-Newton
\citep{brinkman2001,audard2001,pandey2012,didel2024_hr} and Chandra
\citep{ayres2001,nordon2007,nordon2008,huenemoerder2013,bozzo2024}, including the
first-light observation of the former. \citet{bozzo2024} revealed that the line shift
follows the radial velocity change expected for the primary. From the observations
during 1978--2024, the flare frequency is estimated as 57 flares over 142-day total
observing time \citep{didel2024_hr}. In the largest flares, the flux reached
$>5\times10^{-11}~\mathrm{erg}~\mathrm{s^{-1}}~\mathrm{cm^{-2}}$ in the 0.1--12~keV
energy band and the temperature is $\sim 3$ keV \citep{audard2001,perdelwitz2018}.

\subsection{Observation}\label{s2-2}
For the results presented here, we used Resolve \citep{kelley2025,ishisaki2025}, one of
the two scientific instruments onboard XRISM. Resolve is a cryogenic non-dispersive
X-ray spectrometer \citep{mccammon1984} with excellent spectroscopic performance along
with a wide energy coverage of 0.3--12~keV. The energy resolution and gain accuracy are
$\Delta E \sim 4.5$ eV (full width at half maximum, FWHM) and 0.3~eV, respectively in
the Fe K band \citep{porter2024, eckart2024}
. However, the energy band pass is restricted to above 1.7~keV due to the
transmission of the Be window in the instrument gate valve \citep{midooka2021}, which
remains closed on orbit as of today.

The observation (sequence numbers 100011010--100011012) was conducted from 2024 March 06
1:15 to March 10 22:40. The observation was performed both for science and instrument
calibration. As such, there were several additional features of this observation that
enhanced the data for calibration, but had only some minor impacts on the science
observation.

For calibration purposes
, the observation was divided into four parts. Each part was a combination of using one
of the two onboard modulated X-ray calibration sources (MXS1 and MXS3) and one of two
on-board processing modes (normal and ``forced midres''). Briefly, the two MXS sources
\citep{de2018,shipman2024,sawada2024} each can only illuminate about half of the focal
plane array with the instrument gate valve closed, thus each one was used for half of
the two processing modes. The calibration sources (MXS along with the $^{55}$Fe sources
on the filter wheel) were used only during Earth occultation to check the energy scale,
and thus do not affect the observation of HR~1099 directly. The normal mode is for the
standard processing, in which high resolution (HR) graded events are processed with HR
templates onboard \citep{ishisaki2018}. The forced midres mode is for calibration, in
which HR events are processed instead with mid resolution (MR) templates, which only have a quarter of the length of HR templates. A low count rate source such as HR~1099
yields predominantly HR events. However, with the forced mid-res mode, we can check the
energy scale for MR events independent of count rate. The first half of this observation
was performed in normal mode, and the second half in forced midres mode.

The only impacts of this calibration configuration on the science observation were: (1) half of
the observation was in the forced midres mode with an average energy resolution of 4.8
eV at 6 keV, while the other half was in normal mode with an average energy resolution
of 4.5 eV at 6 keV, for high-resolution graded events \citep{porter2024} 
and
(2) there were much more frequent time-dependent energy scale fiducial measurements than
a standard Resolve observation of typically 2--3 measurements per day with
$\sim$30 minutes each. This was done to build up enough statistics from the onboard
calibration sources in both normal and forced midres modes to verify the time-dependent
energy scale correction. 
However, the additional calibration periods
have little effects on the science observation of HR1099 except to contribute to the
verification that the Resolve in-flight calibration process is effective~\citep{porter2024, eckart2024}.

\subsection{Data reduction}\label{s2-3}
The archived data was originally processed using pipeline processing version
03.00.013.009 \citep{doyle2022}. The data was then reprocessed using the
\texttt{xapipeline} ftool included in \texttt{HEASoft} 6.35.1 and \texttt{HEASoft
XRISM}\_\texttt{CalDB11} in the following manner to account for the non-standard
instrument configuration during this observation. Separate gain history files for the
normal and forced-midres parts of the observations were constructed using the gain
fiducial measurements (using on-board $^{55}$Fe sources in the standard method;
\citealt{porter2024}) independently, but in a similar manner, to that implemented in
standard processing. The data in these files allow a per-pixel, time-dependent,
energy-scale calibration to be applied in reprocessing using the \texttt{userghf}
parameter in the \texttt{xapipeline} task.

In addition, the gain curves stored in the calibration database
(\texttt{XRISM\_CalDB11}) were applied such that the MR gain was applied to HR events
for the data acquired in forced midres mode as described above. After screening, the
total on-source time was $\sim$231 ks. Spectra used for analysis were extracted from
cleaned event files with the following standard additional post-processing screening
applied: (1) cuts based on pulse rise time and pixel-to-pixel coincidence
\citep{mochizuki2025} and (2) removal of all events in pixel 27, which is known to
experience sudden, unpredictable gain variations. We analyze only HR grade events
regardless of whether on-board processing was conducted using the HR or MR template,
using the appropriate energy scale conversion and response matrices.

The \texttt{xaarfgen} and \texttt{rslmkrmf} tasks in the \texttt{HEASoft} version 6.35.1
were used to generate the telescope effective area function (ARF) and detector
redistribution matrix file (RMF), respectively. When some event grades are excluded from
analysis, the RMF ordinarily should be scaled by the fraction of event grades
kept. Resolve data, however, contain many low resolution primary (Lp) events
that are induced by cosmic rays, which are followed by multiple false low resolution
secondary (Ls) events by the secondary-pulse detection algorithm in the clipped Lp
event.  Since the rate of real Ls events from HR 1099 is negligible, we removed all Ls
events prior to generating the RMF. As with the gain, HR grade events during the forced
midres mode part of the observation were treated as if they were MR grade events in
generating the RMF since the line spread function reflects the shorter MR grade template
applied to them. The resulting ARF is consistent with that for an on-axis point source.

The non-X-ray background level was checked using the model available
online\footnote{\url{https://heasarc.gsfc.nasa.gov/docs/xrism/analysis/nxb/nxb_spectral_models.html}},
with its normalization determined from the observed spectrum in the 14--17 keV
range. For energy binning of the observed spectra, the optimal binning method
\citep{kaastra2016} was applied, imposing a condition of a minimum of 10 counts per
bin. The spectral fitting was performed to minimize the $\chi^2$ statistics using the
\texttt{xspec} software \citep{arnaud1996}. The uncertainties quoted here are for 90\%
statistical error.

\begin{table*}[]
  \tbl{Best-fit parameters in the thermal plasma model (\S~\ref{s3-2}) and DEM
 analysis (\S~\ref{s4-1-1}).}{%
\begin{tabular}{cccclccc}
\hline
                           & \multicolumn{3}{c}{Quiescent}                                                                                                               &  & \multicolumn{3}{c}{Flare}                                                                                                                     \\ \cline{2-4} \cline{6-8}
Parameters                 & \multicolumn{2}{c}{fiducial 2$kT$}                                                       & DEM                                                       &  & \multicolumn{2}{c}{fiducial 2$kT$}                                                       & DEM                                                         \\ \cline{2-3} \cline{6-7}
                           & low-$kT$                     & high-$kT$                     &                                                                              &  & low-$kT$                    & high-$kT$                     &                                                             \\ \hline
$kT$ (keV)                 & $1.36_{-0.05}^{+0.04}$                 & $3.76_{-0.33}^{+0.27}$                 & \multirow{2}{*}{(figure \ref{f06})}                       &  &   $1.59_{-0.1}^{+0.1}$               & $4.76_{-0.37}^{+0.37}$               & \multirow{2}{*}{(figure \ref{f06})}                          \\
$EM$ ($10^{54}$ cm$^{-3}$) & $2.8_{-0.14}^{+0.12}$                  & $0.51_{-0.07}^{+0.13}$                 &                                                           &  &  $2.8_{-0.22}^{+0.20}$                &   $1.0_{-0.14}^{+0.16}$                &                                                             \\
$Z_{\mathrm{Si}}$\footnotemark[*] & \multicolumn{2}{c}{$0.39_{-0.05}^{+0.05}$}                                      & $0.39_{-0.05}^{+0.05}$                             &  & \multicolumn{2}{c}{$0.51_{-0.12}^{+0.13}$}                                      & $0.35_{-0.08}^{+0.08}$                                      \\
$Z_{\mathrm{S}}$\footnotemark[*] & \multicolumn{2}{c}{$0.32_{-0.03}^{+0.03}$}                                      & $0.32_{-0.02}^{+0.02}$                             &  & \multicolumn{2}{c}{$0.41_{-0.06}^{+0.06}$}                                      & $0.32_{-0.04}^{+0.04}$                                      \\
$Z_{\mathrm{Ar}}$\footnotemark[*] & \multicolumn{2}{c}{$0.70_{-0.07}^{+0.07}$}                                      & $0.64_{-0.06}^{+0.06}$                             &  & \multicolumn{2}{c}{$0.74_{-0.12}^{+0.12}$}                                      & $0.63_{-0.10}^{+0.10}$                                       \\
$Z_{\mathrm{Ca}}$\footnotemark[*] & \multicolumn{2}{c}{$0.55_{-0.08}^{+0.08}$}                                      & $0.49_{-0.07}^{+0.07}$                             &  & \multicolumn{2}{c}{$1.13_{-0.16}^{+0.17}$}                                      & $1.00_{-0.14}^{+0.14}$                                      \\
$Z_{\mathrm{Fe}}$\footnotemark[*] & \multicolumn{2}{c}{$0.28_{-0.02}^{+0.02}$}                                      & $0.28_{-0.02}^{+0.02}$                             &  & \multicolumn{2}{c}{$0.44_{-0.03}^{+0.03}$}                                      & $0.43_{-0.03}^{+0.03}$                                      \\
$Z_{\mathrm{Ni}}$\footnotemark[*] & \multicolumn{2}{c}{$0.45_{-0.32}^{+0.33}$}                                      & $0.56_{-0.35}^{+0.36}$                             &  & \multicolumn{2}{c}{$0.59_{-0.38}^{+0.38}$}                                      & $0.64_{-0.38}^{+0.38}$                                      \\ \hline
Reduced $\chi^2$ (dof)            & \multicolumn{2}{c}{1.04 (1937)}                    & 1.05 (1941)                                                                     &  & \multicolumn{2}{c}{0.98 (1694)} & 1.06 (1698)                                 \\ \hline
\end{tabular}
}\label{t02}
\begin{tabnote}
\footnotemark[$*$] The elemental abundances are relative to those of \citet{wilms2000}.\\
\end{tabnote}
\end{table*}

\begin{figure*}[!hbtp]
 \begin{center}
 \includegraphics[width=0.681\linewidth]{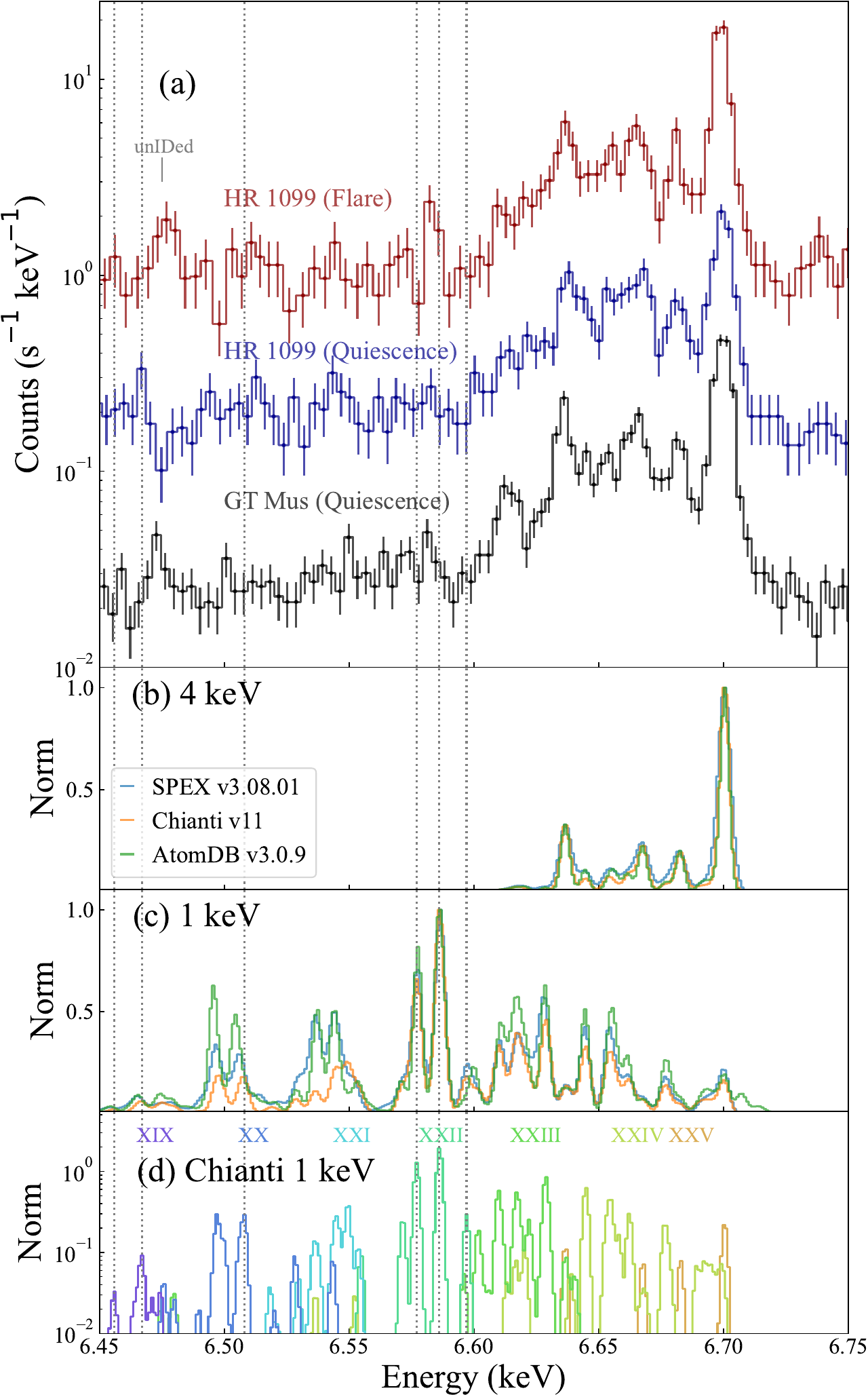}
 \includegraphics[width=0.309\linewidth]{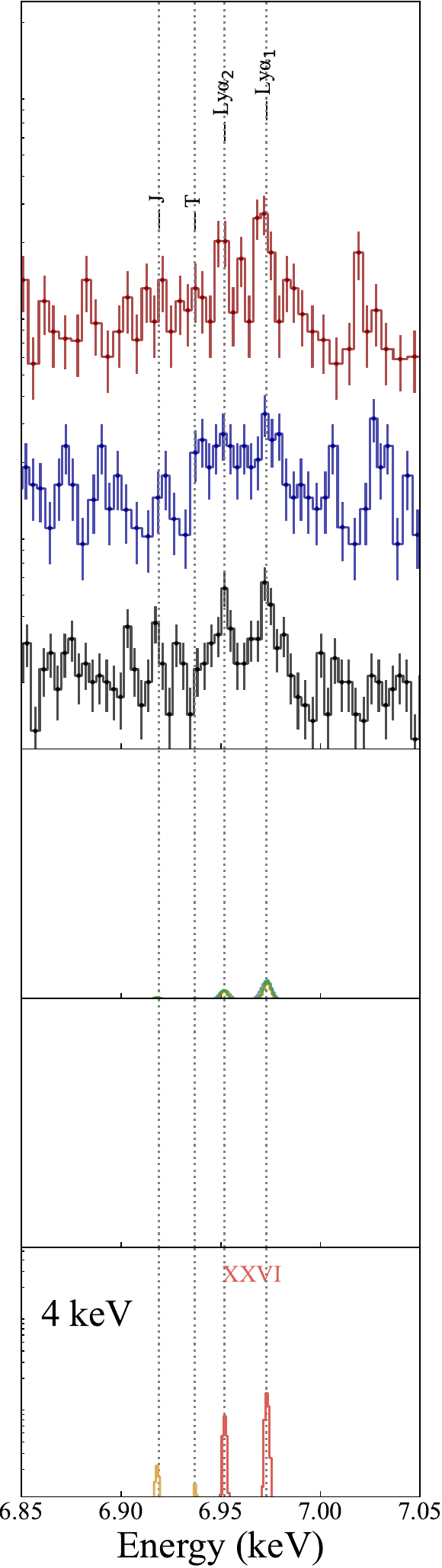}
 \end{center}
 \caption{(a) Close-up view of the Resolve spectra in the Fe
 \emissiontype{XIX}--\emissiontype{XXVI} (left) and Fe \emissiontype{XXVI} (right)
 K-shell complexes of HR~1099 in flare (red) and quiescence (blue) compared to that of
 GT Mus in quiescence (black; \citealt{kurihara2025}). The intensity is scaled by
 $\times$20 for the red, $\times$10 for the blue, and $\times$1 for the black spectra to
 facilitate comparison. The centroid energy of the lines used for the DEM construction
 is indicated by vertical dashed lines.
 (b, c) Synthesized spectra with three different plasma codes: \texttt{SPEX} version
 3.08.01 (blue), \texttt{Chianti} version 11 (orange), and \texttt{AtomDB} version 3.0.9
 (green) for a thermal plasma of (b) 4~keV or (c) 1~keV. The spectra are smoothed with a
 Gaussian of an FWHM$=4$ eV, normalized at the strongest line of Fe \emissiontype{XXVI}
 Ly$\alpha_1$ for $kT=4$ keV and Fe \emissiontype{XXII} line 6.586 keV for $kT=1$ keV.
 (d) Synthesized spectra (1~eV binning) with \texttt{Chianti} version 11 decomposed for
 each ion.
 Alt text: A four-panel figure of the Fe K complex of the (a) Resolve spectra,
 (b,c) model spectra from \texttt{SPEX} version 3.08.01, \texttt{Chianti} version 11,
 and \texttt{AtomDB} version 3.0.9, and (d) model spectra for individual ions.}
 \label{f04}
\end{figure*}

\section{Analysis}\label{s3}
\subsection{Light curve and spectra}\label{s3-1}
The X-ray light curve for the entire observation is shown in figure \ref{f01} (a). On
top of the baseline count rate ($0.25 \pm 0.07$ s$^{-1}$) that varies slowly with a
small amplitude, several sudden increases were found. 
The largest increase peaked at approximately 180~ks from the start of the observation, reaching 1.6~s$^{-1}$, and lasted until about 220~ks.
There are several smaller events at $\sim$150, $\sim$200, and
$\sim$260~ks. It happened to be that the second of the four parts of the observation
contained most of the largest event and two smaller events. We thus use this duration to
extract the flare spectrum and all the others for the quiescent spectrum. The total
radiation energy of the flare is estimated to be $\approx 10^{34}$ erg in the
1.7--10~keV range.

The Resolve spectra for the quiescent and flare phases are presented in
figure~\ref{f02}. Both spectra exhibit a rich set of emission lines superimposed on the
continuum emission, including He-like and H-like transitions from elements such as Si,
S, Ar, Ca, Cr, Fe, and Ni. The flare spectrum shows a substantial increase in both
continuum and line fluxes, reflecting enhanced high-temperature plasma emission during
the flare. It is confirmed that the background spectrum is almost negligible, except for
the Au L line at $\sim 9.7$ keV, thus the background spectrum was not subtracted in the
spectrum fitting shown below.

\begin{table*}
  \tbl{Line list for the phenomenological model (\S~\ref{s3-3}) and the results of DEM
 analysis (\S~\ref{s4-1-1}).}{%
 \begin{tabular}{ccllcccccccc}
\toprule
Label\footnotemark[$*$]   & Ion               & Lower                                           & Upper                             & $E$ \footnotemark[$\dag$] &\multicolumn{3}{c}{Quiescent} & &\multicolumn{3}{c}{Flare}\\ \cline{6-8} \cline{10-12}
 & & & & (keV) & Det.\footnotemark[$\ddag$] & Norm ($10^{-6}$)\footnotemark[$\S$] & Ratio\footnotemark[$\|$]  & & Det.\footnotemark[$\ddag$] & Norm ($10^{-6}$)\footnotemark[$\S$] & Ratio\footnotemark[$\|$] \\
\midrule
\multicolumn{12}{c}{(Fe \emissiontype{XIX}--\emissiontype{XXII} K-shell band)}\\
                          & Fe\emissiontype{XIX}  & 1s$^{2}$.2s$^{2}$.2p$^{4}$~$^3$P$_1$    & 1s.2s$^{2}$.2p$^{5}$~$^3$P$_2$                         & 6.4558  &U & $<0.90$              & --- & &U & $<1.2$                & --- \\
                          & Fe\emissiontype{XIX}  & 1s$^{2}$.2s$^{2}$.2p$^{4}$~$^3$P$_2$    & 1s.2s$^{2}$.2p$^{5}$~$^3$P$_2$                         & 6.4669  &Y & $0.12_{-0.1}^{+0.3}$ & $0.9_{-0.7}^{+0.7}$ & &U& $<2.0$ & --- \\
                          & Fe\emissiontype{XX}   & 2s$^{2}$.2p$^{3}$~$^4$S$_{3/2}$         & 1s.2s$^{2}$.2p$^{4}$~$^4$P$_{5/2}$                     & 6.4971  &U & $<0.56$              & --- & &N & ---                   & --- \\
                          & Fe\emissiontype{XX}   & 2s$^{2}$.2p$^{3}$~$^4$S$_{3/2}$         & 1s.2s$^{2}$.2p$^{4}$~$^4$P$_{1/2}$                     & 6.5080  &Y & $0.41_{-0.4}^{+0.4}$ & $1.3_{-1.0}^{+1.0}$ & &U& $<1.9$ & --- \\
                          & Fe\emissiontype{XXII} & 1s$^{2}$.2s$^{2}$.2p~$^2$P$_{3/2}$      & 1s.2s$^{2}$.2p$^{2}$~$^2$D$_{5/2}$                     & 6.5771  &Y & $0.49_{-0.4}^{+0.4}$ & $0.84_{-0.6}^{+0.6}$ & &U& $<0.97$ & --- \\
                          & Fe\emissiontype{XXII} & 1s$^{2}$.2s$^{2}$.2p~$^2$P$_{1/2}$      & 1s.2s$^{2}$.2p$^{2}$~$^2$D$_{3/2}$                     & 6.5861  &U & $<0.55$              & --- & &Y & $2.2_{-1.3}^{+1.3}$  &  $0.90_{-0.6}^{+0.6}$ \\
                          & Fe\emissiontype{XXII} & 1s$^{2}$.2s$^{2}$.2p~$^2$P$_{3/2}$      & 1s.2s$^{2}$.2p$^{2}$~$^2$S$_{1/2}$                     & 6.5970  &Y & $0.46_{-0.4}^{+0.4}$ & $3.0_{-2.5}^{+2.5}$ & &U& $<1.5$ & --- \\
\midrule
\multicolumn{12}{c}{(Fe \emissiontype{XXIII}--\emissiontype{XXV} K-shell band)}\\
$E12$                     & Fe\emissiontype{XXIII}& 1s$^{2}$.2s.2p~$^3$P$_2$                & 1s.2s.2p$^{2}$~$^3$D$_3$                               & 6.6097  &Y & $0.90_{-0.5}^{+0.5}$ & $12_{-6.9}^{+6.4}$ & &Y& $2.1_{-1.5}^{+1.4}$    & $1.6_{-1.1}^{+1.1}$ \\
$u$\footnotemark[$\#$] & Fe\emissiontype{XXIV} & 1s$^{2}$.2s~$^2$S$_{1/2}$               & 1s.2s($^3$S).2p~$^4$P$_{3/2}$                             & 6.6167  &N & ---                  & --- & &U & $<2.7$ & --- \\
$e$\footnotemark[$\#$] & Fe\emissiontype{XXIV} & 1s$^{2}$.2p~$^2$P$_{3/2}$               & 1s.2p$^{2}$($^3$P)~$^4$P$_{5/2}$                          & 6.6203  &Y & $1.4_{-1.0}^{+0.5}$  & $2.9_{-2.0}^{+1.1}$ & &Y&  $3.0_{-2.6}^{+1.6}$  & $2.6_{-2.3}^{+1.4}$ \\
$E3$ ($\beta$)            & Fe\emissiontype{XXIII}& 1s$^{2}$.2s$^{2}$~$^1$S$_0$             & 1s.2s$^{2}$.2p~$^1$P$_1$                               & 6.6288  &Y & $0.95_{-0.6}^{+0.6}$ & $0.90_{-0.6}^{+0.5}$ & &Y& $3.9_{-1.8}^{+1.8}$  & $1.8_{-0.8}^{+0.8}$ \\
$z$                       & Fe\emissiontype{XXV}  & 1s$^{2}$~$^1$S$_0$                      & 1s.2s~$^3$S$_1$                                        & 6.6366  &Y & $3.7_{-0.8}^{+0.8}$  & $1.1_{-0.2}^{+0.2}$ & &Y&  $10.5_{-2.3}^{+2.4}$ & $0.85_{-0.2}^{+0.2}$ \\
$j$                       & Fe\emissiontype{XXIV} & 1s$^{2}$.2p~$^2$P$_{3/2}$               & 1s.2p$^{2}$($^1$D)~$^2$D$_{5/2}$                       & 6.6447  &Y & $2.4_{-0.8}^{+0.7}$  & $0.83_{-0.3}^{+0.2}$ & &Y&  $5.1_{-2.0}^{+1.9}$ & $0.76_{-0.3}^{+0.3}$ \\
$r$                       & Fe\emissiontype{XXIV} & 1s$^{2}$.2s~$^2$S$_{1/2}$               & 1s.2s($^1$S).2p~$^2$P$_{1/2}$                          & 6.6529  &U & $<2.4$               & --- & &U    & $<7.8$ & --- \\
$k$                       & Fe\emissiontype{XXIV} & 1s$^{2}$.2p~$^2$P$_{1/2}$               & 1s.2p$^{2}$($^1$D)~$^2$D$_{3/2}$                       & 6.6547  &U & $<3.6$               & --- & &U& $<9.0$  & --- \\
$a$                       & Fe\emissiontype{XXIV} & 1s$^{2}$.2p~$^2$P$_{3/2}$               & 1s.2p$^{2}$($^3$P)~$^2$P$_{3/2}$                       & 6.6579  &N & ---                  & --- & &N & --- & ---\\
$q$                       & Fe\emissiontype{XXIV} & 1s$^{2}$.2s~$^2$S$_{1/2}$               & 1s.2s($^3$S).2p~$^2$P$_{3/2}$                          & 6.6622  &Y & $2.2_{-2.0}^{+0.9}$  & $1.2_{-1.1}^{+0.5}$ & &Y& $5.6_{-2.6}^{+2.5}$ & $1.2_{-0.6}^{+0.5}$\\
$y$                       & Fe\emissiontype{XXV}  & 1s$^{2}$~$^1$S$_0$                      & 1s.2p~$^3$P$_1$                                        & 6.6676  &Y & $3.7_{-0.9}^{+1.3}$  & $1.6_{-0.4}^{+0.6}$ & &Y& $9.3_{-2.4}^{+2.6}$ & $1.1_{-0.3}^{+0.3}$\\
$t$\footnotemark[$\#$] & Fe\emissiontype{XXIV} & 1s$^{2}$.2s~$^2$S$_{1/2}$               & 1s.2s($^3$S).2p~$^2$P$_{1/2}$                             & 6.6762  &Y & $0.90_{-0.4}^{+0.6}$ & $0.98_{-0.5}^{+0.7}$ & &Y& $1.8_{-1.7}^{+1.6}$  & $0.79_{-0.78}^{+0.7}$\\
$x$                       & Fe\emissiontype{XXV}  & 1s$^{2}$~$^1$S$_0$                      & 1s.2p~$^3$P$_2$                                        & 6.6827  &Y & $2.5_{-0.7}^{+0.9}$  & $1.1_{-0.3}^{+0.4}$ & &Y& $7.6_{-2.4}^{+2.2}$   & $0.97_{-0.3}^{+0.3}$\\
$d13$                     & Fe\emissiontype{XXIV} & 1s$^{2}$.3p~$^2$P$_{3/2}$               & 1s.2p($^1$P).3p~$^2$D$_{5/2}$                          & 6.6892  &N & ---                  & --- & &Y & $0.32_{-0.3}^{+3.1}$ & --- \\
$d15$                     & Fe\emissiontype{XXIV} & 1s$^{2}$.3p~$^2$P$_{1/2}$               & 1s.2p($^1$P).3p~$^2$D$_{3/2}$                          & 6.6917  &U & $<1.9$               & --- & &Y & $2.1_{-2.0}^{+2.1}$ & --- \\
$w$                       & Fe\emissiontype{XXV}  & 1s$^{2}$~$^1$S$_0$                      & 1s.2p~$^1$P$_1$                                        & 6.7004  &Y & $9.6_{-1.2}^{+1.3}$  & $0.94_{-0.1}^{+0.1}$ & &Y& $41.1_{-4.1}^{+4.1}$ & $1.1_{-0.1}^{+0.1}$\\
\midrule
\multicolumn{12}{c}{(Fe \emissiontype{XXVI} K-shell band)}\\
$J$                       & Fe\emissiontype{XXV}  & 1s.2p~$^1$P$_1$                         & 2p$^{2}$~$^1$D$_2$                                     & 6.9188  &N & --- & --- & &U & $<1.8$ &  ---\\
$T$                       & Fe\emissiontype{XXV}  & 1s.2s~$^1$S$_0$                         & 2s.2p~$^1$P$_1$                                        & 6.9373  &U & $<0.45$ & --- & &U & $<2.3$ &  ---\\
Ly$\alpha_2$              & Fe\emissiontype{XXVI} & 1s~$^2$S$_{1/2}$                        & 2p~$^2$P$_{1/2}$                                       & 6.9521  &Y & $1.3_{-0.6}^{+0.6}$ & $1.3_{-0.6}^{+0.6}$ & &Y& $3.3_{-1.5}^{+1.5}$ & $0.99_{-0.4}^{+0.5}$\\
Ly$\alpha_1$              & Fe\emissiontype{XXVI} & 1s~$^2$S$_{1/2}$                        & 2p~$^2$P$_{3/2}$                                       & 6.9732  &Y & $1.4_{-0.6}^{+0.6}$ & $0.89_{-0.4}^{+0.4}$ & &Y& $5.4_{-1.7}^{+1.8}$ & $0.96_{-0.3}^{+0.3}$\\
\bottomrule
\end{tabular}
}\label{t01}
\begin{tabnote}
\footnotemark[$*$] Notations are from \citet{beiersdorfer1993} for Fe \emissiontype{XXIII}
 lines, \citet{gabriel1972} for Fe \emissiontype{XXIV} $n=2\rightarrow1$ lines, and
 Safronova's \citep{bitter1984} for Fe \emissiontype{XXV} satellite lines.\\
\footnotemark[$\dag$] Line energy from \texttt{Chianti} version 11 \citep{dere1997,dufresne2024}.\\
\footnotemark[$\ddag$] Whether the line is detected with a 90\% significance: ``Y'
 (detected), ``U'' (not detected, but an upper limit is estimated to be above the NXB
 level), ``N'' (not detected, and an upper limit is lower than the NXB level). \\ 
\footnotemark[$\S$] The best-fit Gaussian norm values of the phenomenological fit. \\ 
\footnotemark[$\|$] The ratio of the input from best-fit phenomenological modeling (\S~\ref{s3-3}) over the DEM analysis output (\S~\ref{s4-1-1}). Those from
 lines not used in the DEM analysis are indicated with ``---''.\\
\footnotemark[$\#$] Lines with possible blending with others commented in
 \citet{beiersdorfer1993}: $u$ with (Fe \emissiontype{XXIV} $v$, Fe \emissiontype{XXIII}
 $E8$, and Fe \emissiontype{XXIII} $E9$), $e$ with (Fe \emissiontype{XXIII} $E6$ and Fe
 \emissiontype{XXIII} $E7$), and $t$ with Fe \emissiontype{XXIV} $m$.\\ 
\end{tabnote}
\end{table*}

\subsection{Broadband spectral modeling}\label{s3-2}
We first describe the entire spectrum using an emission model from the collisionally
ionized plasma at a thermal equilibrium and at the optically thin and low-density limits
(\texttt{APEC} model; \citealt{smith2001}). The model parameters are the plasma
temperature ($kT$), emission measure ($EM$), and elemental abundance relative to solar
($Z$). The abundance table in \citet{wilms2000} was used. The model was attenuated by
the photo-electric absorption model \citep{wilms2000} for the interstellar extinction
with the H-equivalent column density ($N_{\mathrm{H}}$), which is set to be $10^{18}$
cm$^{-2}$ by following previous studies on this target
\citep{huenemoerder2013,didel2024_hr}.

For both the quiescent and flare spectra, we first fitted with a single-temperature
model with $kT \sim 2$~keV, which left significant residuals at both low- and
high-energy ends. We then fitted with a two-temperature model with $kT <2$~keV and
$kT>2$~keV with a common 
$Z$ but with a different $EM$. This
yielded a reasonable result (table \ref{t02}). This is a traditional approach to be
compared to the DEM analysis presented later, and we call this fiducial 2$kT$ model
hereafter.

\subsection{Fe K-shell line complex modeling}\label{s3-3}
\begin{figure*}[!hbtp]
 \begin{center}
 \includegraphics[width=0.99\linewidth]{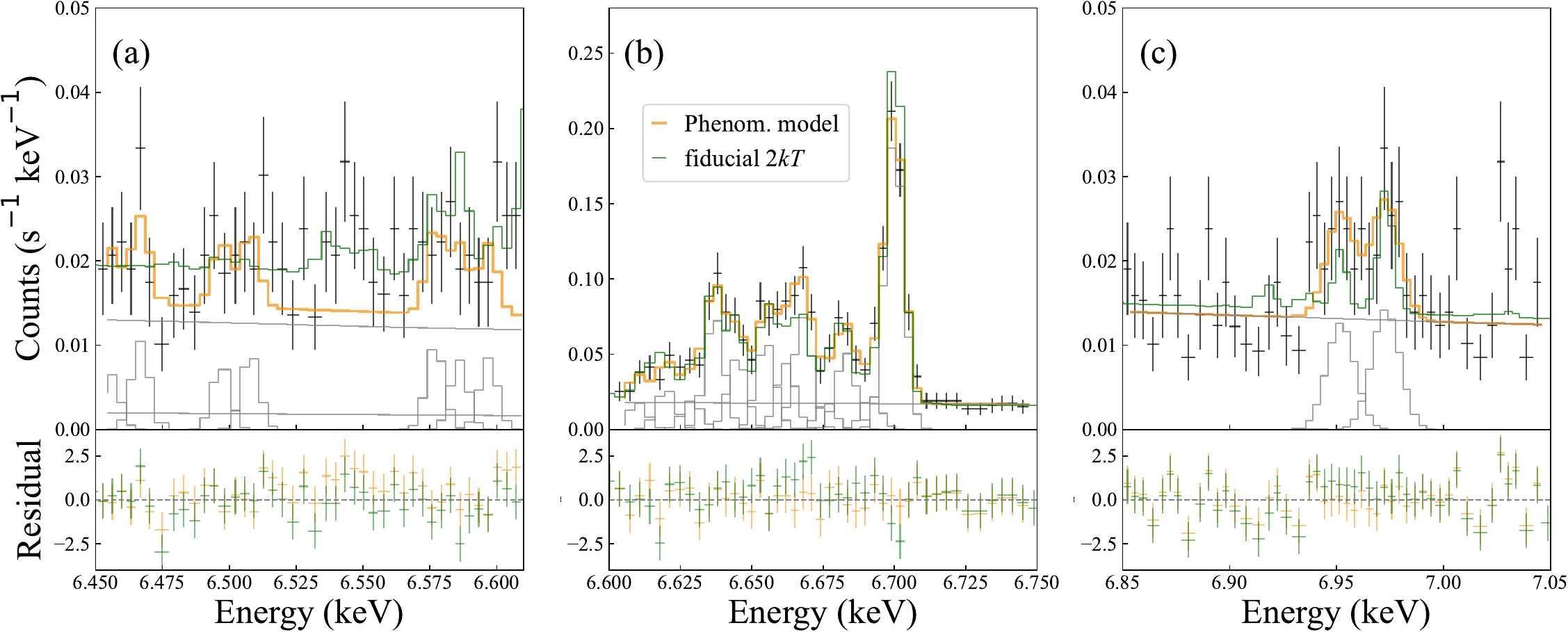}
 \includegraphics[width=0.99\linewidth]{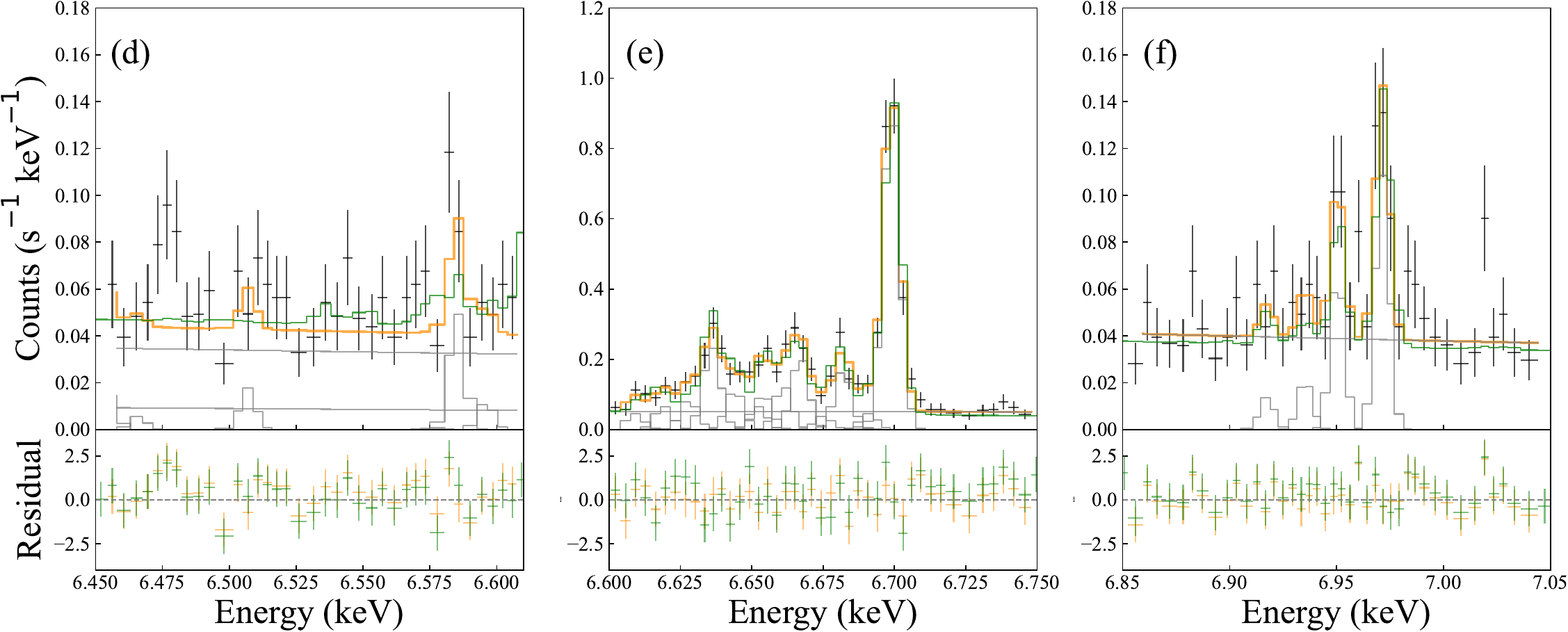}
 \end{center}
 \caption{Close-up views of the Resolve spectrum (black) of the Fe K-shell line
 complexes, the best-fit model with the fiducial 2$kT$ model (green) and
 phenomenological model (orange), and their residuals. The upper panels (a)--(c) are for the quiescent
 spectrum, while the lower panels (d)--(f) are for the flare spectrum.
 Alt text: A pair of three-panel figures showing close-up views of the Resolve
 spectra in the Fe K-shell line complexes and the best-fit spectral models for the
 quiescent and flaring phases, respectively.  }
 \label{f05}
\end{figure*}

\begin{table*}[]
  \tbl{Best-fit parameters in the phenomenological model (\S~\ref{s3-3}).}{%
\begin{tabular}{lccccccc}
\hline
\multirow{2}{*}{Parameter} & \multicolumn{3}{c}{Quiescent}                                               &  & \multicolumn{3}{c}{Flare}                                                   \\ \cline{2-4} \cline{6-8} 
                           & Fe \emissiontype{XIX}--\emissiontype{XXII} K-shell         & Fe \emissiontype{XXIII}--\emissiontype{XXV} K-shell         & Fe \emissiontype{XXVI} K-shell         &  & Fe \emissiontype{XIX}--\emissiontype{XXII} K-shell         & Fe \emissiontype{XXIII}--\emissiontype{XXV} K-shell         & Fe \emissiontype{XXVI} K-shell \\ \hline
Redshift (eV)              & Fixed\footnotemark[$*$]                   & $0.0^{+0.5}_{-0.5}$     & Fixed\footnotemark[$*$]                   &  & Fixed\footnotemark[$*$]                  & $1.0^{+0.3}_{-0.3}$      & $1.6^{+1.0}_{-1.0}$     \\
Broadening (eV)              & Fixed\footnotemark[$*$]                   & $2.5^{+0.7}_{-0.6}$     & $6.2^{+3.8}_{-2.4}$     &  & Fixed\footnotemark[$*$]                  & $2.1^{+0.4}_{-0.4}$      & $2.4^{+1.4}_{-1.3}$     \\
Reduced $\chi^2$ (dof)     & 1.41 (35) & 0.67 (26) & 1.18 (40) &  & 1.24 (30) & 0.90 (24) & 0.82 (36) \\ \hline
\end{tabular}
}\label{t03}
\begin{tabnote}
\footnotemark[$*$] The value is fixed to the best-fit value in the high He$\alpha$ band in the same phase.\\
\end{tabnote}
\end{table*}

\subsubsection{Inspection}\label{s3-3-1}
We start by inspecting the spectrum (figure~\ref{f04}a) for the Fe
\emissiontype{XIX}--\emissiontype{XXV} K-shell complex (left) and Fe \emissiontype{XXVI} K-shell
complexes (right). The HR~1099 spectra in the flare and the quiescence as well as the GT
Mus spectrum in the quiescence are compared. GT Mus is an RS CVn type binary and is the
only other stellar source observed so far with Resolve \citep{kurihara2025}. We
divide the two complexes into three bands: (1) Fe
\emissiontype{XIX}--\emissiontype{XXII} K-shell band (6.450--6.603~keV), (2) Fe
\emissiontype{XXIII}--\emissiontype{XXV} K-shell band (6.603--6.750~keV), and (3) Fe
\emissiontype{XXVI} K-shell band (6.850--7.050~keV). A list of lines in each band is
provided in table~\ref{t01} with their conventional labels, if any, ion, lower and upper
levels, and line center energy. These lines carry rich information for the DEM
construction.

In the Fe \emissiontype{XXVI} K-shell band, the main lines of Fe \emissiontype{XXVI} as
well as the dielectronic recombination (DR) satellite lines of Fe \emissiontype{XXV} are
included. The Fe \emissiontype{XXVI} Ly$\alpha_1$ and Ly$\alpha_2$ (6.973 and 6.952~keV)
lines probe the electron distribution around $\sim$5~keV. They can be seen in both
phases and appears enhanced during the flaring phase (figure~\ref{f04}). The clear
separation of Ly$\alpha_1$ and Ly$\alpha_2$ and the major DR lines $J$ and $T$ requires
the spectral resolution $R \gtrsim 500$ in this band, which is unique to
Resolve among X-ray spectrometers for stellar sources.

In the Fe \emissiontype{XXIII}--\emissiontype{XXV} K-shell band, the main lines of Fe
\emissiontype{XXV} ($w$: 6.700, $x$: 6.683, $y$: 6.668, and $z$: 6.637 keV) as well as
the satellite lines of Fe \emissiontype{XXIII}--\emissiontype{XXIV} ($q$: 6.662, $j$:
6.645 keV, etc) are included. The satellite to main line ratio is a robust estimator of
the electron temperature due to their different line formation processes as was
demonstrated for GT Mus with Resolve \citep{kurihara2025}. For example, the
upper level of the $w$ line is predominantly populated via the electron collision
excitation of He-like Fe, that of the $q$ line is via the electron collision excitation
of Li-like Fe, and that of the $j$ line is via the DR of He-like Fe
\citep{mochizuki2025b}.  A quick comparison of the nearby $q$ and $y$ lines indicates
the change of the Fe charge state distribution, hence the DEM distribution.

In the Fe \emissiontype{XIX}--\emissiontype{XXII} K-shell band, the Fe K-shell lines of
lower ionization stages are included. The access to these lines is also unique to
Resolve for stellar sources. The lines below Fe\emissiontype{XXI} have been
rarely investigated even in the Sun \citep{doschek1981}.  Since some lines are clearly
detected and identified in Resolve spectra, we try to use them to constrain the
temperature around $\sim$1~keV (figure~\ref{f03}).

\subsubsection{Fitting}\label{s3-3-2}
We fitted the spectrum in each band using a phenomenological model, which consists of
lines listed in table~\ref{t01} upon the continuum emission. The individual lines are
represented with a Gaussian model with a fixed energy based on \texttt{Chianti} version
11 \citep{dere1997,dufresne2024} and a normalization as a free parameter. All lines are
collectively fitted for the parameters of the energy shift and the broadening. The continuum emission is
represented by the \texttt{NLAPEC} model \citep{smith2001}, which is equivalent to the
\texttt{APEC} model for thermal plasma emission but without emission lines having
emissivities larger than $10^{-20}$~cm$^3$ s$^{-1}$. The best-fit model is illustrated
in figure~\ref{f05}, while the best-fit parameters are tabulated in tables~\ref{t01} and
\ref{t03}. We also checked the possibility of the plasma deviating from a collisionally
ionized equilibrium using the ratios of the Fe \emissiontype{XXVI} Ly$\alpha_1$, Fe
\emissiontype{XXV} $w$ and $J$, Fe \emissiontype{XXIV} $j$ and $q$ \citep{kurihara2025}, which yielded no conclusive evidence for any deviations.

A redshift of 50$\pm15$~km~s$^{-1}$ was observed during the flare (table~\ref{t03}). The sign of the observed radial velocity is
consistent with that expected from the orbital motion of the primary star
\citep{bozzo2024} at around phase~$=0.1$, when the flare reached its peak, although the
measured velocity shift is somewhat larger. The difference may be attributable to
additional kinematic effects associated with the flare, such as chromospheric
evaporation and/or coronal mass ejection \citep{inoue2024}, as well as to the relative
motion of the satellite with respect to the system. The limited statistics do not allow
us to investigate this further, and will be left out of the scope of this study.

\section{Discussion}\label{s4}
\subsection{DEM distribution}\label{s4-1}
\subsubsection{Choice of lines}\label{s4-1-1}
We first consider which lines to use for the DEM construction. Those in the Fe
\emissiontype{XXIII}--\emissiontype{XXV} K-shell and Fe \emissiontype{XXVI} K-shell
bands are relatively straight-forward and well established in ground experiments
\citep{beiersdorfer1993,decaux1997,rudolph2013} and solar observations (e.g.,
\citealt{phillips2004i,watanabe2024}) and are used with Resolve
\citep{kurihara2025}. In contrast, those in Fe \emissiontype{XIX}--\emissiontype{XXII}
K-shell band are less established even in ground experiments and solar observations and
are new with Resolve.

We compared three different plasma codes ---\texttt{Chianti} version 11
\citep{dere1997,dufresne2024}, \texttt{AtomDB} version 3.0.9
\citep{smith2001,foster2012} and \texttt{SPEX} version 3.08.01
\citep{kaastra1996,kaastra_2024}--- in figure~\ref{f04} (b). Their synthesized spectra
exhibit larger discrepancies in the Fe \emissiontype{XIX}--\emissiontype{XXII} K-shell
band than the other two bands, suggesting the level of systematics. 
All three codes were compared to the Hitomi observation of the Perseus cluster using the X-ray microcalorimeter \citep{aharonian2018}.
Among the three
codes, \texttt{Chianti} is the most well-verified with solar observations, but still has significant limitations, especially for the lower charge states. The model of Fe \emissiontype{XVIII}--\emissiontype{XXIII} was
updated in version 9, although comparisons with solar observations were made only for Fe
\emissiontype{XXII} and the higher charge states \citep{dere2019}. Two results
illustrate the limitation. One is that some strong lines of the lower charge states (Fe
\emissiontype{XIX} and \emissiontype{XX}) were observed during solar flares
\citep{phillips2004,feldman2006}, but they do not match with \texttt{Chianti} in line
center or strength. The other is that a clear but unidentified line feature at
$\sim$6.475~keV was observed in the Resolve data of HR~1099 during the
flare, which might also be present in the GT Mus data (figure~\ref{f04}a). The line is too strong to be a
Fe\emissiontype{XIX}--\emissiontype{XXI} line, or an Fe \emissiontype{XXIII} line that exists
around 6.479 keV (figure~\ref{f04}c). We also confirmed that the line does not match
with known lines of  He- or H-like ions of Cr and Mn including their Rydberg series (\citealt{hell2025} and references therein).

Considering all these, we decided to use all detected lines in the Fe
\emissiontype{XXVI} K-shell and Fe \emissiontype{XXIII}--\emissiontype{XXV} K-shell
bands except for the Fe \emissiontype{XXIV} $d15$ in the flaring phases (as it belongs
to the unresolved infinite series of DR transitions onto levels with $n \geq 3$), and
some strong and relatively isolated detected lines in the Fe
\emissiontype{XIX}--\emissiontype{XXII} K-shell band for the Fe \emissiontype{XXII},
\emissiontype{XX}, and \emissiontype{XIX} that have corresponding structures found
by eye in Resolve spectra. The Fe \emissiontype{XXI} lines were not used, as
nearby lines from other charge states, especially Fe \emissiontype{XXIV} $o$ and $p$ are
present with possible large contributions from high-$kT$ components.

\subsubsection{DEM construction}\label{s4-1-2}
\begin{figure*}[!hbtp]
 \begin{center}
 \includegraphics[width=0.79\linewidth]{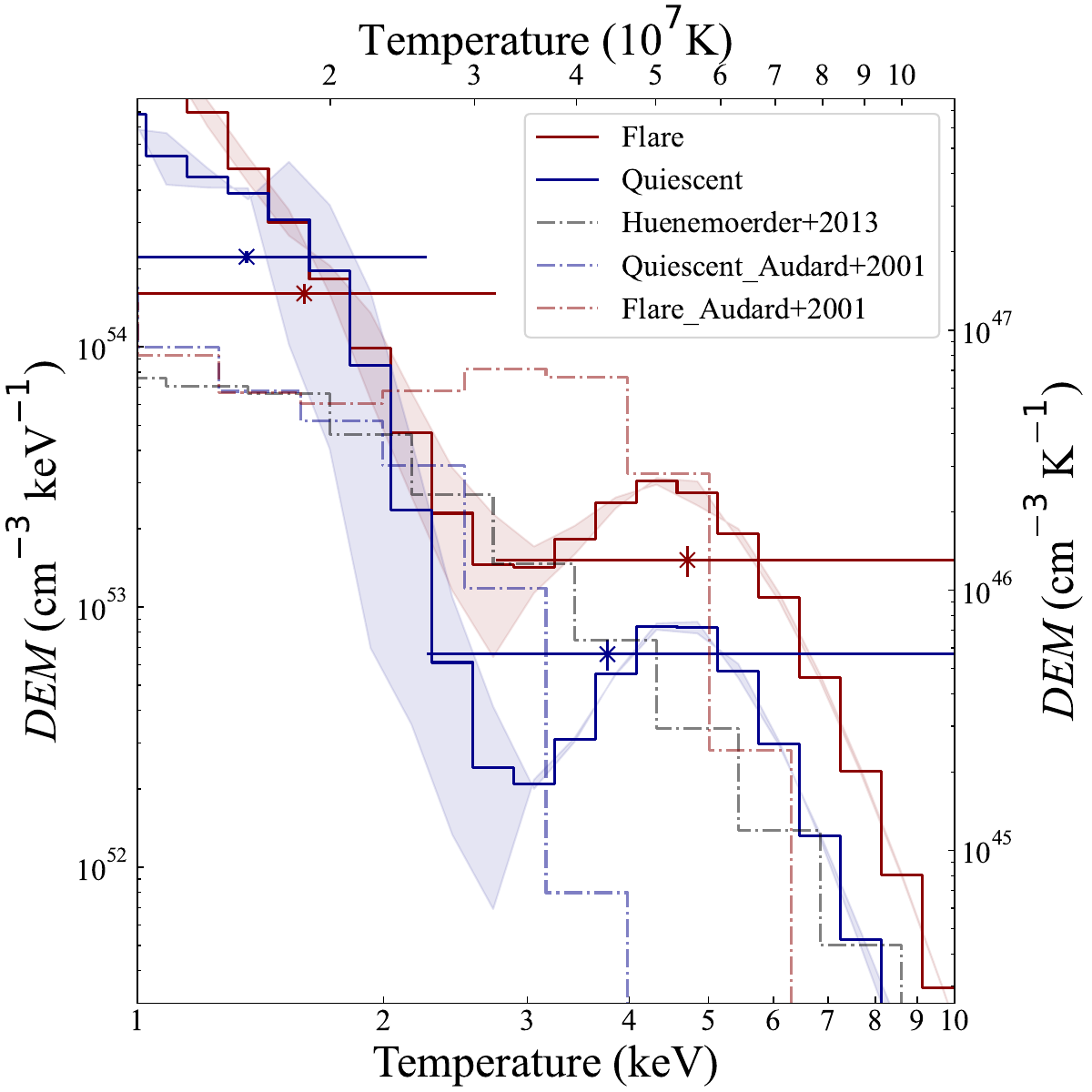}
 \end{center}
 \caption{DEM (histogram with errors) derived from the series of Fe K-shell lines for the
 quiescent (blue) and flaring (red) phases. Discrete data points indicate the DEM derived
 from the fiducial 2$kT$ model, and their horizontal error bars are divided by the geometric means of the two best-fit temperatures at each phase. A DEM derived in the previous studies are also shown:  \citet{huenemoerder2013} using Chandra in gray,  \citet{audard2001} using XMM-Newton for quiescence in dashed blue and flare in dashed red, respectively. 
 Alt text: A figure showing DEM for the quiescent and the flaring phases as well as one
 from the previous studies. 
 } 
 \label{f06}
\end{figure*}

There are two major approaches conventionally used to model the temperature structure of
the stellar corona in X-ray spectroscopy. One is to describe the global spectra
using multiple discrete temperature components, which is often used in medium-resolution
CCD spectra. This method is advantageous for the use of both the line and continuum
emission, which is useful in particular when the Bremsstrahlung cutoff lies within the
spectrum range. The other is to account for individual line intensities with pre-calculated
contribution functions, which is often used in high-resolution grating spectra both in
solar and stellar studies. This allows us to constrain parametric or even
non-parametric models of the DEM distribution directly from the fitting.
In this paper, we adopt the latter approach, which was made possible by the increased
number of resolved Fe K-shell lines with Resolve.

Several techniques have been developed for constructing DEM distributions in solar
studies (\citealt{del2018} for review). Typically, these represent a compromise between allowing significant structure
in the DEM to attempt to maximize the number of line intensities that can be
reproduced, and adding smoothing to the DEM to try to capture a more physically realistic
temperature distribution. After an initial survey of the Resolve dataset using
a variety of techniques, we decided to adopt the method introduced by \citet{warren2005}
for solar EUV spectra. In this technique, the DEM distribution is parameterized with a
spline function of the temperature. The parameters are constrained through $\chi^2$
fitting of the line intensities using the contribution functions calculated with
\texttt{Chianti}. This approach provides improved flexibility in reproducing the steep
decline in the DEM distribution at the highest temperature end, a behavior that, as we
show below, is visible in the HR~1099 data, and other multi-temperature fitting
approaches often struggle to capture.

We first derived the shape of the DEM distribution using only the Fe K lines selected in \S~\ref{s4-1-1}. We then determined the scaling factor of the DEM distribution by referring back to the observed spectra in the 1.7--10.0 keV band for the purpose of decoupling the degeneracy between the normalization of the DEM distribution and the Fe abundance. The DEM model spectra were synthesized by summing \texttt{APEC} models with the corresponding temperatures and emission measures. After adjusting the abundances of the other elements and the overall normalization, the DEM model yielded fit statistics comparable to those of the fiducial 2$kT$ models, as shown in table \ref{t02}.

\subsubsection{Result}\label{s4-1-3}
Figure~\ref{f06} presents the DEM distributions derived with uncertainties separately
for the quiescent and flare phases. We assessed the success of our analysis by examining
the reproduction rate of each line, defined as the ratio of the synthesized line
intensity to the observed line intensity, for each DEM distribution
(table~\ref{t01}). This is expected to be unity, which is satisfied for most lines
within the statistical uncertainty. A few exceptions are Fe \emissiontype{XXIII} $E12$, Fe \emissiontype{XXV} $y$, and $w$ during the quiescent phase, which may be
attributable to blending with neighboring lines.  Nevertheless, a sufficiently large
number of lines were successfully used to constrain the DEM distribution over the
1--10~keV temperature range. 

The DEM exhibits a bimodal distribution both for the quiescent and flare phases peaking
at $\sim$1 and $\sim$5~keV. The hotter component of the two increased during the flare,
while the colder component remained almost unchanged. Although the temperature range is
shifted higher in HR 1099, the bimodal distribution is often observed when the Sun is viewed as a
star (e.g., \citealt{caspi2014} and references therein).
The valley in the DEM distribution at $\sim$3~keV stems from the observed weakness of
the Fe\emissiontype{XXIV} line intensities compared to the other Fe lines; they would
have been stronger if the DEM distribution was flat across $\sim$3~keV. Such a
constraint is difficult to obtain in the traditional multi-$kT$ modeling, as the
$\chi^{2}$ local minimum can be easily obtained without accounting for weak lines.

\subsubsection{Comparison}\label{s4-1-4}
We compared the DEM distribution of HR~1099 in previous results using HETG during
quiescence \citep{huenemoerder2013} and in a flare \citep{nordon2008} and that using RGS
in both flare and quiescence \citep{audard2001}. The distributions in
\citet{huenemoerder2013} and \citet{audard2001} are shown in figure~\ref{f06}. The DEM distributions are
somewhat different from one another. 
While the distributions in the two HETG and RGS quiescent phase results 
are roughly single-peaked, that in the Resolve results shows a double peak. 
In the solar
case, multiple-peaked EM distributions have been shown to arise from a number of
uncertainties \citep{lanzafame2002}, and we list several possible additional reasons for
the difference here. 
The first is that the DEM depends on how it is reconstructed. For instance, the RGS result in the quiescent phase is based on the traditional multi-$kT$ approach, while the two HETG results are derived from individual line fitting based on
contribution functions. 
The second is that RGS and HETG lack sensitivity to the lines of
Fe\emissiontype{XXV}, Fe\emissiontype{XXVI}, or both. 
Third, the flare in the HETG observation \citep{nordon2008} is much weaker than
the one detected with Resolve or RGS; the count rate increased only by 10\%. In
fact, a small increase at 260~ks from the start in the Resolve data
(figure~\ref{f01}), which we recognized as a part of quiescence, is larger than the
flare in \citet{nordon2008}. The quiescent phase in \citet{huenemoerder2013} has some
variation within a factor of 2 in the light curve \citep{ayres2001}. In large stellar
flares in other stars observed with grating spectrometers, the bimodal DEM distribution
and the increase only of the hotter component appears to be the norm, as was seen in
$\sigma$ Gem \citep{nordon2006,huenemoerder2013}, and AD Leo \citep{van2003}.

\subsection{Abundance}\label{s4-2}
\begin{figure}[!hbtp]
 \begin{center}
 \includegraphics[width=0.99\linewidth]{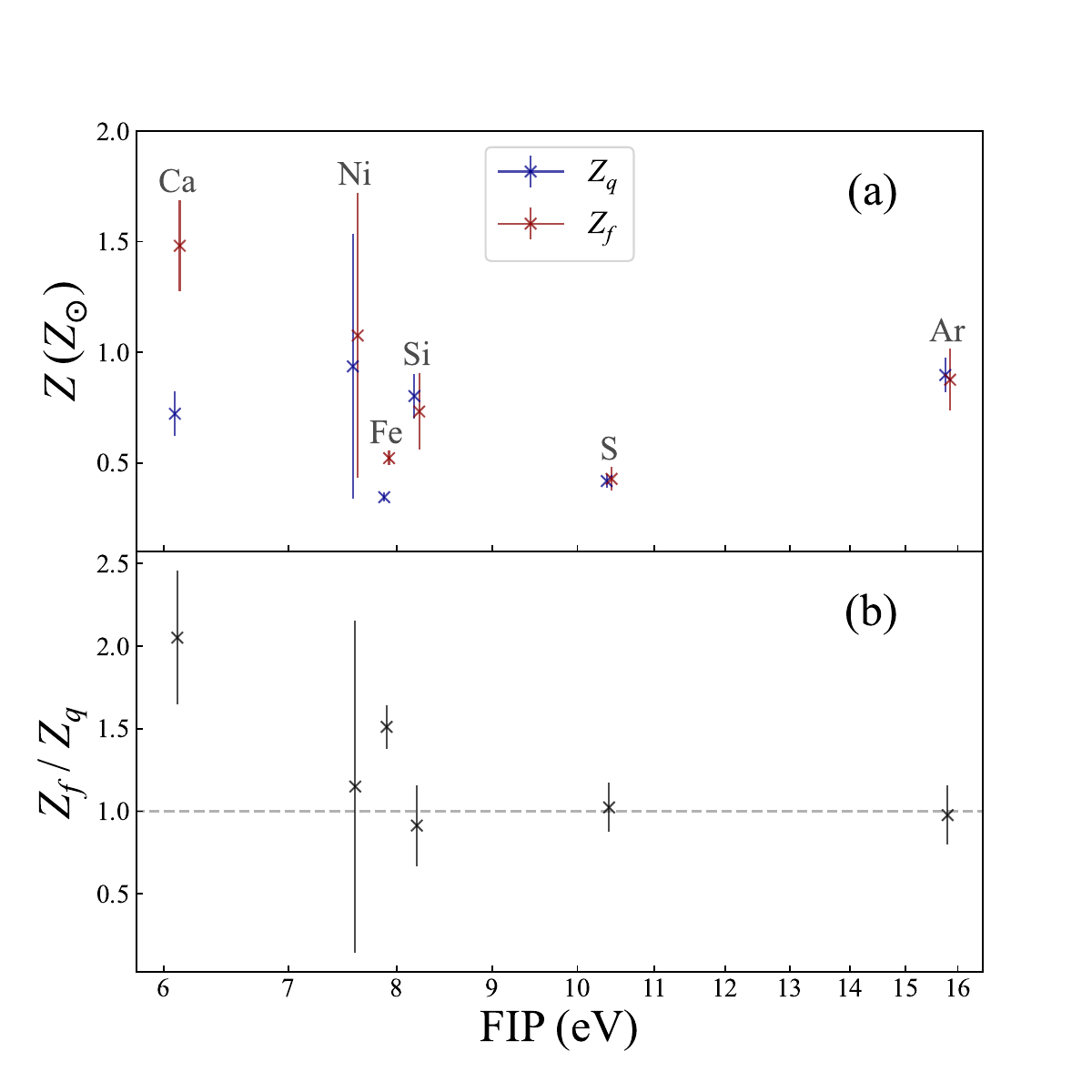}
 \end{center}
 \caption{(a) Derived elemental abundance of the quiescent (blue) and the flare (red)
 phase. The abundance values relative to those of \citet{wilms2000} are plotted against the
 FIP. (b) Ratio between the quiescent and flare abundances.
 Alt text: A two-panel figure showing the measured abundances. 
 } 
 \label{f08}
\end{figure}

Based on the DEM distribution derived solely from the Fe K-shell lines, we estimated the
abundances of the other elements.
While the relative shape of the DEM was fixed, allowing the vertical scaling to vary, we fitted the abundances of Si, S, Ar, Ca, Fe, and Ni relative to H, together with the DEM normalization, to the 1.7–10keV spectra.
In practice, we use the continuum level, which includes free-free, free-bound, and two-photon emission (with a dominant contribution from hydrogen free-free), primarily as a proxy for the H abundance.
The results are summarized in table \ref{t02}, while the abundance as a function of the
FIP is depicted in figure~\ref{f08} (a). The reference values of FIP are retrieved from
NIST Atomic Spectra Database \citep{NIST_ASD}. Since we do not know the photospheric
abundances of stars such as HR~1099, it is common to interpret their spectra using solar
photospheric abundances. During the quiescent phase, the observed abundance pattern of
Fe, Si, S, and Ar is consistent with previous results for HR~1099 (see figure 7 in
\citealt{didel2024_hr}). Overall, the elemental abundances are sub-solar photospheric,
though the photospheric abundances of HR~1099 may differ from the solar values.

We thus focus on the relative abundances in the flare phase compared to the quiescent
phase within our dataset (figure~\ref{f08}b).  During the flare, the abundances of Ca
and Fe increased clearly by a factor of $\sim2$ for Ca and $\sim50$\% for Fe, while those of the other elements
remained unchanged. This selective enhancement can be understood as follows.
As \citet{nordon2008} suggested in their stellar flare studies and as
confirmed also in solar flares (e.g., \citealt{warren2014}), chromospheric evaporation
\citep{neupert1968,gudel2002} can transport plasma with different chemical compositions
than those in the coronae into the flare loop. The fact that HR~1099 nominally exhibits
the i-FIP effect and only the two elements with the lowest FIP in our sample showed
enhancement during the flare provides support for the evaporation scenario.  In solar
studies, Si is often regarded as a low-FIP element, but no significant variation was
detected here. It is unclear whether this reflects the diverse nature of the FIP effect
or an observational bias, given that the He$\alpha$ features of Si lie close to the
lower end of the energy band, unlike those of the other elements.

One proposed explanation for the FIP fractionation is the ponderomotive force, which
acts only on charged particles \citep{laming2015,laming2021}. In this context, Ca is
particularly interesting: it has a low first ionization potential (6.1~eV) and a second
ionization potential (11.9~eV), making it the only abundant element whose second
ionization potential lies below the first ionization potential of H
(13.6~eV). Observational evidence for the relative enhancement of elements with lower
FIP than Fe has so far been scarce in stellar cases (e.g., Mg in AT Mic;
\citealt{raassen2003at}), despite many reports of abundance variations during flares in
general (e.g., \citealt{favata1999,nordon2007,nordon2008,liefke2010}).  This first
detection of Ca enhancement presented here demonstrates the complementary
spectroscopic capabilities of XRISM to the grating missions, highlighting its value in
probing the diverse manifestations of the FIP effect in stars.

To investigate this further, we should make assessments similar to figure~\ref{f08} (b)
in more elements, including larger stellar flares. When a giant solar flare is resolved
temporally, or spatially, or both, or an in-situ measurement is employed, an increase of
the low-FIP elements can be observed at some times, or in some locations, such as the
initial phase of the flare, at the loop top, or in the slow wind
\citep{katsuda2020,suzuki2025,to2024,brooks2022}. With better photon statistics,
time-resolved spectroscopy will enable comparisons with the Sun as well as the reduction
of observational bias due to the lack of spatial resolution. 
Furthermore, giant flares may allow the detection of very low-FIP elements of alkali
metals such as Na (5.1~eV) and K (4.3~eV) with high-resolution X-ray spectroscopy,
despite their small abundances \citep{huenemoerder2013}. As is demonstrated by previous
results with grating spectrometers and the present result with the microcalorimeter
spectrometer, we now have instruments and diagnostic tools to make this
possible. Investments of long telescope times to capture flares are awaited in the
future.

\section{Summary}\label{s5}
We presented the result of the Resolve observation of an RS CVn star HR1099. A
stellar flare was observed with an X-ray microcalorimeter for the first time. In
addition to the Fe He$\alpha$ and Ly$\alpha$ complexes modeled in our previous study of
GT Mus with Resolve \citep{kurihara2025}, lines from relatively lower charge
states of Fe \emissiontype{XIX}--\emissiontype{XXIII} were utilized for the
reconstruction of the DEM distribution, effectively extending Fe K-shell diagnostics pioneered
by solar missions such as P78-1 \citep{doschek1979}, the Solar Maximum Mission (SMM;
\citealt{culhane1981}), and Hinotori \citep{tanaka1982}. The DEM was derived using a
single element, Fe, for both the quiescent and flaring phases, covering a temperature
range from $\sim$1--10 keV. We found that the DEM distribution is bimodal, and only the
hotter component increased during the flare. Based on the DEM distribution thus
obtained, the elemental abundances of Si, S, Ar, Ca, Fe, and Ni were derived. 
The observed increase of abundances only at the lower end of the FIP range (Ca, Fe) constitutes a clear example of the abundance change during stellar flares as a function of the FIP.
Along with previous results using grating spectrometers, we demonstrated
the possibility of using X-ray high-resolution spectroscopy to investigate the chemical
fractionation processes in the context of coronal heating and flare physics.

\begin{ack}
 This article is published in PASJ as open access, published by OUP (\url{https://doi.org/10.1093/pasj/psaf124}).
\medskip

 We thank the anonymous reviewer for their constructive comments that helped improve the manuscript.
 We gratefully acknowledge insightful discussion with Tetsuya Watanabe (NAOJ) on
 high-resolution X-ray spectroscopy, and with Andy S. H. To (ESTEC) on solar
 observations of the abundance variation during flares. 
 David P. Huenemoerder (MIT) provided
 the Chandra result for figure~\ref{f06}.
 This research has made use of data and/or software provided by the High Energy
 Astrophysics Science Archive Research Center (HEASARC), which is a service of the
 Astrophysics Science Division at NASA/GSFC. \texttt{Chianti} is a collaborative project
 involving George Mason University, the University of Michigan (USA), University of
 Cambridge (UK) and NASA Goddard Space Flight Center (USA).
 This research was supported by the grant of Joint Research by the National Institutes
 of Natural Sciences (NINS program No OML032402). The material is based upon work
 supported by NASA under award number 80GSFC21M0002.
 M.\,K. was supported by JSPS Grant-in-Aid for JSPS fellows 25KJ0926 and the IGPEES, WINGS
 Program, the University of Tokyo, and the ACME project, which has received funding from the European Union's Horizon Europe Research and Innovation programme under Grant Agreement No 101131928.
 E.\,B. acknowledges support from NASA grants 80NSSC20K0733, 80NSSC24K1148, and
 80NSSC24K1774. Part of this work was performed under the auspices of the
 U.S. Department of Energy by Lawrence Livermore National Laboratory under Contract
 DE-AC52-07NA27344.
 The work of D.H.B. was performed under contract to the Naval Research Laboratory and was funded by the NASA Hinode program.
\end{ack}

\bibliographystyle{aa}
\bibliography{ms1}
\end{document}